\documentclass[journal,final,10pt,twocolumn]{IEEEtran}
\normalsize
\usepackage{amsmath, amsthm, amssymb}
\usepackage{epsfig}
\usepackage{mathtools}
\usepackage{latexsym}   
\usepackage{epstopdf}
\usepackage{multirow}
\usepackage{cite}
\usepackage{comment}
\usepackage{algorithm}
\usepackage{caption}
\usepackage{subcaption}
\usepackage{algpseudocode}
\usepackage{float}
\usepackage{graphicx}
        \usepackage{setspace}

\usepackage{xcolor}

\begin{document}

\setlength{\abovedisplayskip}{3pt}
\setlength{\belowdisplayskip}{3pt}

\title{Latency Reduction for Mobile Edge
Computing in HetNets by Uplink and Downlink Decoupled
Access}

\author{Ali Al-Shuwaili and Ahmed Lawey }

\maketitle

\begin{abstract}
Achieving an end-to-end low-latency for computations offloading, in Mobile Edge Computing (MEC) systems, is still a critical design problem. This is because the offloading of computational tasks via the MEC servers entails the use of uplink (UL) and downlink (DL) radio links that are usually assumed to be coupled to a single base station (BS). However, for heterogeneous networks, a new architectural paradigm whereby UL and DL are not associated with the same BS is proposed  and seen to provide gains in network throughput due to the improved UL performance. Motivated by such gains, and by using typical results from stochastic geometry, we formulate the offloading latency for the MEC-based scheme with decoupled UL/DL association, or decoupled access, and compare its performance to the conventional coupled access scheme. Despite the backhaul delay necessary for the communication between the two serving BSs in UL and DL, the offloading scheme with decoupled access is still capable of providing a fairly lower offloading latency compared to the conventional offloading scheme with coupled access.
\end{abstract}


\let\thefootnote\relax\footnotetext{A. Al-Shuwaili is with the College of Engineering, University of Information Technology and Communications, Baghdad, Iraq (e-mail: ali.najdi@uoitc.edu.iq).

A. Lawey is with the School of Electronic and Electrical Engineering, University of Leeds, United Kingdom (e-mail: a.q.lawey@leeds.ac.uk).

}

\section{Introduction}

Leveraging the  computational capabilities  of  the nearby Base Stations (BSs), known as Mobile Edge Computing (MEC), seems to be unavoidable technique to cope with the computing and battery capacity limitations of mobile devices \cite{cloudlet}.  However, excessive delay  might be experienced during the communications between Mobile Users (MUs) and MEC, or cloudlet, servers due to variable channel and interference conditions. Given that offloading entails the use of UpLink (UL) and DownLink (DL) radio links, many lines of work have demonstrated that it is possible to design an energy- and latency-efficient MEC systems by, for example, performing a joint optimization of the UL/DL allocation of communication and computational resources \cite{ali}.

To meet the stringent latency requirements for delay-sensitive applications like medical or AR/VR applications, feasible  offloading time needs to be in order of  milliseconds \cite{ali}. Such critical latency values are potentially limited by the MU-BS association type, or access type, employed by the network and by the number of offloading users in each tier as these two factors determine the resulting interference level in UL and DL links. The standard structure of current wireless networks constraints users to associate to the same BS in both uplink and downlink. In DL, the MUs first associate  to the BS that provide the highest average power, and then use the same BS for UL transmission \cite{why,dude}. This \textit{coupled} association scheme is efficient for traditional cellular network where single type of BSs are regularly deployed and have identical radio capabilities. However, with the emergence of heterogeneous networks (HetNets) where different types of BSs, like macro, femto or pico, are coexisted in multi-tier set-up,  an emerging paradigm in 5G systems that is shown to improve the capacity of HetNets is  by treating UL and DL as separate network connections, i.e, \emph{decoupled} access \cite{why,dude}.

  Motivated by the impact of access type, i.e., coupled or decoupled on the offloading latency and also the importance of taking into account backhaul capacity limitations, since backhaul is well understood to be often the bottleneck in dense HetNets \cite{ali}, we formulate novel  expressions, using stochastic geometry tools, for the latency in MEC-based offloading scheme while decoupled access is assumed for UL and DL connections and analyze its performance. 
  
  The limited literature on computations offloading with decoupled access includes papers \cite {ref1,ref2,ref3}. The work in \cite{ref1} proposes MEC-aware association rule and compare its performance, via Monte Carlo simulation, to the traditional coupled access scheme, taking the scenario of task offloading in the UL as an example. Reference \cite{ref2} introduces the decoupled access in Fog Radio Access Networks (F-RANs) and jointly optimizes
the user access and offloading decisions to minimize energy consumption by using the reinforcement learning algorithm.  The study in \cite{ref3} addresses
the problem of energy-efficient user association in Cloud Radio Access Networks (C-RANs)
via joint optimization of MU association and the BSs muting. Section II introduces the system model  and also the formulation of the offloading latency expressions while numerical results are provided in Section III. Concluding remarks are finally provided in Section IV. 

\section{System Model}

A heterogeneous mobile edge computing network that consists of a two-tier deployment of Macro cell Base Stations (MBSs) and Small cell Base Stations (SBSs) is considered. Both tiers operate on the same frequency band and using Frequency Division Duplex (FDD) \cite{why}.  The locations of BSs in the $k$th tier, with $k\in \{M,S\}$, are modeled according to a two-dimensional homogeneous Poisson Point Process (PPP) $\Phi_k$ with density  $\lambda_k$. The transmission powers of all BSs in the same tier are assumed to be identical and are denoted as $P_k$ with $k\in\{M,S\}$. The locations of the MUs in the network are also modeled according to a homogeneous PPP $\Phi_u$ with density $\lambda_u$ that is independent of the BS locations $\Phi_M$ and $\Phi_S$. We also assume that the MUs in the same BS use an OFDMA-like orthogonal multiple access scheme, such that there is exactly one user per cell that is scheduled on the same time-frequency resource each with transmit power of $P_u$. Signals in both UL and DL are assumed to experience path loss with path loss exponent $\alpha > 2$. Each receiver has a constant noise power  of $\sigma^2$.

A local computing server, or ``cloudlet'', is  directly connected to each BS in both tiers (see Fig. 1).  We generally assume that the MBSs' cloudlets have a higher computation capacity as compared to the SBSs' cloudlets \cite{cloudlet}. Denoting as $F_k$ the cloudlet computation capacity  in CPU cycles per seconds for the BSs in tier $k$, we then have $F_M \geq F_S$. Also, a SBS is connected to the closest MBS via a finite capacity backhaul link. The capacity in bits per seconds of the backhaul link   that connects an SBS to an MBS is denoted as $C^{bh}$. 

MU-BS association policy can be either coupled or decoupled. With the conventional coupled access, each MU is assigned in both UL and DL to the BS that offers maximal received power in the DL. Instead, with decoupled access, the MU is associated in the DL to the BS from which it receives the maximal power while, for the UL, it is associated to the BS that receives its signal with highest average power \cite{petar,dude}.

The overall set of MUs is scheduled for offloading, in which, each MU wishes to run an application that is defined by the number $V$ of CPU cycles necessary to process one request from that application, by the number $B^I$ of input bits necessary to offload the computations of one request to the cloudlet processor, and by the number $B^O$ of output bits encoding the result of the computation (see, e.g., \cite{ali}). To offload an application, the MU first transmits the $B^I$ input bits  to its associated MBS or SBS; then a cloudlet executes the $V$ CPU cycles; and finally the $B^O$ output bits   are sent back to the MU. As it will be discussed, the execution of the application can take place at the cloudlet attached to the BS to which the MU is connected in the UL or to the BS to which the MU is connected in the DL. In fact, the UL and DL BSs need not to be the same in the presence of a decoupled access policy. In such cases, backhaul transmission is necessary for the  communications between the the UL and DL BSs. The association model for both coupled and decoupled access is discussed next. Throughout, the analysis will assume the existence of the typical MU, i.e., an MU that is located at the origin \cite{petar, andre_dl}.

\begin{figure}%
    \centering
    \subfloat[\centering Coupled access]{{\includegraphics[width=2cm]{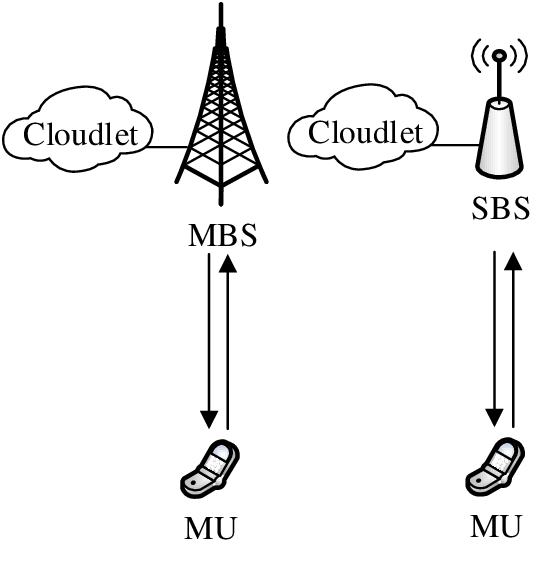} }}%
    \qquad
    \subfloat[\centering Decoupled access]{{\includegraphics[width=2.5cm]{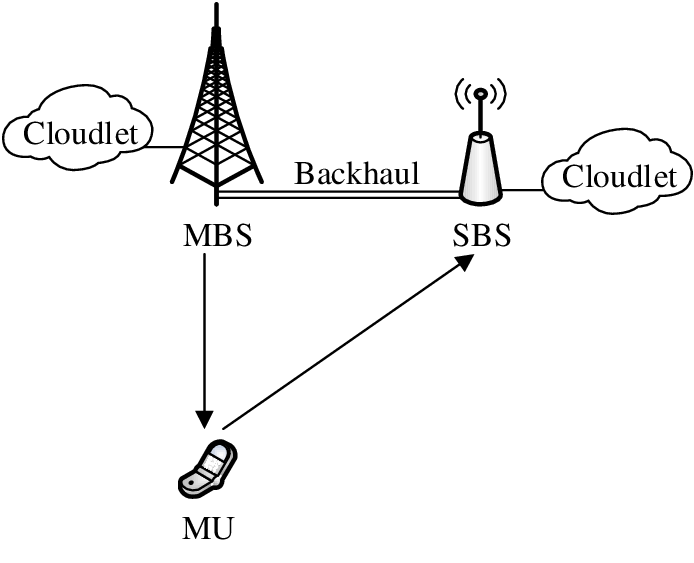} }}%
    \caption{System model: Mobile users (MUs) offload the execution of their applications to a cloudlet processor in two-tier heterogeneous networks. Note that coupled, or DL-based, association implies  case (a) while decoupled, or UL-based,  association implies (a) and (b) cases.}%
    \label{fig:example}%
\end{figure}

\subsection{Association Model}

Let $\|x_k^0\|$ be the distance from the typical MU to the nearest $k$BS with $k \in \{M,S\}$, we then formulate the association rule as
\begin{equation} \label{assoc}
 \underset{ k \in \Phi_M \cup \Phi_S} {\rm argmax} \; T_k \|x_k^0\|^{-\alpha} , 
\end{equation}
where $T_k$ is a parameter that specifies the access or association type in UL and DL as explained next.

\textit{(1) Coupled access:}  It is a DL-based association policy where, by setting $T_k=P_k$ with $k \in \{M,S\}$ in (\ref{assoc}), the MU is associated to the BS from which it receives the highest average power. The same BS will also be   the serving BS for the uplink connection. As a result, with the considered two-tier HetNet, the coupled association  can lead to two possible association cases: {\textit{(i)} Case 1: UL base station = DL base station = SBS} and \textit{(ii)} Case 2: UL base station = DL base station = MBS. We denote by $A_{kl}^{D}$ the association probability when the MU is associated to a $k$BS in UL and a $l$BS in DL with $k=l \in \{M,S\}$ and the association is DL-based or coupled. The corresponding association probabilities for these two cases can be obtained from the general expression in \cite[Lemma 1]{andre_dl} as:

\emph{(i) Case 1: UL base station = DL base station = SBS:} 
\begin{equation}
 A_{SS}^{D}= \dfrac{\lambda_S}{\lambda_S+ (\frac{P_M}{P_S})^{2/\alpha} \lambda_M};  
\end{equation}

\emph{(ii) Case 2: UL base station = DL base station = MBS:} 
\begin{equation}
A_{MM}^{D}= \dfrac{\lambda_M(\frac{P_M}{P_S})^{2/\alpha} }{\lambda_S+ (\frac{P_M}{P_S})^{2/\alpha} \lambda_M}.
\end{equation}

\textit{(2) Decoupled access:} It is an UL-based association policy whereby, by setting $T_k=1$ with $k \in \{M,S\}$ in  (\ref{assoc}), the MU is associated to the BS to which it transmits with the highest average power. The concept of decoupled access implies, based on (1), that MU can select two different base stations, each one corresponds to different network connection, i.e., UL and DL. For this case, the association process can lead to one of the  four following possible association cases. We also  denote by $A_{kl}^{U}$ the association probability when the MU is associated to a $k$BS in UL and a $l$BS in DL with $k,l \in \{M,S\}$ and the association is UL-based or decoupled. Following the procedures in \cite{petar}, the association probabilities for these four cases are given as follows:

\emph{(i) Case 1: UL base station = DL base station = SBS:} 
\begin{equation}
 A_{SS}^{U}=\dfrac{\lambda_S}{\lambda_S+ (\frac{P_M}{P_S})^{2/\alpha} \lambda_M};  
\end{equation}

\emph{(ii) Case 2: UL base station = DL base station = MBS:}
\begin{equation}
A_{MM}^{U}= \frac{\lambda_M}{\lambda_M+\lambda_S};
\end{equation} 

\emph{(iii) Case 3: UL base station = SBS and DL base station = MBS:} 
\begin{equation}
A_{SM}^{U}= \frac{\lambda_S}{\lambda_S+\lambda_M} - \dfrac{\lambda_S}{\lambda_S+ (\frac{P_M}{P_S})^{2/\alpha} \lambda_M};  
\end{equation}

\emph{(iv) Case 4: 	UL base station = MBS and DL base station = SBS:}
\begin{equation}
A_{MS}^{U} = 0.
\end{equation}

 The total offloading latency depends on the type of association which will be discussed in Section II-D after formulating the communication and computation times for coupled and decoupled access in Section II-B and Section II-C, respectively.
\subsection{Communication and Computation Model with Coupled Access}
The offloading latency consists  of  the  time $L^{ul}$
required  for  the  MU  to  send  the
input bits to its serving BS in the uplink; the time $L^{\rm exe}$ necessary for  the  cloudlet  to  process  the  instructions;  the time $L^{bh}$ for exchanging information between BSs in different tiers; and the time $L^{dl}$ to send the result
back  to  the  MU  in  the  downlink (see  Fig.  1).  We  can  hence write the total offloading latency for a typical MU with decoupled access as
\begin{equation} \label{delay}
L=L^{ul}+L^{\rm exe} + L^{bh} + L^{dl}.
\end{equation}
It is noted here that the offloading latency for the coupled access scheme is a special case of (\ref{delay}) which is obtained with $L^{bh}=0$. In the rest of this section, we derive an expression for each latency term in (\ref{delay}) under coupled association. The latency analysis for decoupled association is deferred to Section II-C.

\subsubsection{Uplink transmission} The average rate, in bits/s, for transmitting the input bits of a typical MU in the uplink using DL-based or coupled association is given by
\begin{equation}
R^{ul}_{D,k}(\gamma^{ul}
_k)=\frac{W^{ul}}{N_{D,k} } \log_2(1+\gamma^{ul}_k)  \mathrm{P}_{D,k}^{ul}(\gamma^{ul}_k),
\end{equation}
where $\gamma^{ul}_k$ is the target SINR threshold for MU connected to the $k$th tier in the uplink; $W^{ul}$ is the uplink bandwidth in Hz; $N_{D,k}$ is the average number of uplink MUs associated with a BS in the $k$th tier using DL-based association and is given by $N_{D,k}=\lambda_u A_{kl}^{D} /\lambda_k$ where $A_{kl}^{D}$ is the association probability as given in (2) and (3) with $k=l=S$ for SBS tier and $k=l=M$ for MBS tier, respectively; and  $\mathrm{P}_{D,k}^{ul}(\gamma^{ul}_k)$ is the coverage probability in the UL of HetNet, i.e., the probability that the instantaneous UL SINR is larger than or equal to the corresponding thresholds $\gamma^{ul}_k$ when a typical MU is associated with $k$th tier via coupled association. Throughout, for notation simplicity, we consider identical per-tier SINR thresholds, i.e., $\gamma_k^{ul}=\gamma^{ul}$ for all $k$. The coverage probability is derived in \cite{petar} as 
\begin{equation}
\mathrm{P}_{D,k}^{ul}(\gamma^{ul})= \int_{0}^{\infty} e^{-\frac{\gamma^{ul} \sigma^2 x^{\alpha}}{P_u}} e^{-\pi \tilde \lambda_u \psi(\gamma^{ul},\alpha) x^2 } f_{{X}_k}^{D}(x) dx,
\end{equation}
where $\tilde \lambda_u = p \lambda_u$ with $p$ being the thinning probability given by $p=(\lambda_M+\lambda_S)/\lambda_u$ and $f_{{X}_k}^{D}(x)$ is the Probability Density Function (PDF) of the distance between a typical MU and its serving BS using DL-based association which reads \cite{petar}
\begin{equation}
f_{{X}_k}^{D}(x)=\frac{2 \pi \lambda_k}{A_{kl}^{D}}x e^{- \left(\lambda_k+\lambda_j(\frac{P_j}{P_k})^{(2/\alpha)}\right) \pi x^2}
\end{equation}
and, for $\alpha=4$, $\psi(\gamma^{ul},\alpha)= \sqrt{\gamma^{ul}} \arctan{\sqrt{ \gamma^{ul}}}$ \cite{andre_dl}. The time, in seconds, necessary to complete the uplink transmissions to the $k$th tier in coupled access is defined as
\begin{equation}
L^{ul}_{D,k}(\gamma^{ul})=\frac{B^I}{R_{D,k}^{\rm ul}(\gamma^{ul})}.
\end{equation}

\subsubsection{Downlink transmission}
The downlink rate function, in bits/s, depends on the target SINR threshold level in the DL $\gamma^{dl}_k$ according to the relation
\begin{equation}
R^{ dl}_{D,k}(\gamma^{dl}_k)=\frac{W^{dl}} {N_{D,k}} \log_2(1+\gamma^{dl}_k) \mathrm{P}_{D,k}^{dl}(\gamma_k^{dl}),
\end{equation}
with $W^{dl}$ being the DL bandwidth and $\mathrm{P}_{D,k}^{dl}(\gamma^{dl}_k)$ is the  probability of coverage in the DL, i.e., the probability that the instantaneous SINR is larger than or equal to the corresponding thresholds $\gamma^{dl}_k$ for a typical MU in the DL. Assuming identical per-tier SIR thresholds ($\gamma_k^{dl}=\gamma^{dl}$ for all $k$), the DL coverage probability is derived in \cite{andre_dl} as
\begin{equation}
\mathrm{P}_{D,k}^{dl}(\gamma^{dl})= \int_{0}^{\infty} e^{-\frac{\gamma^{dl} \sigma^2 x^{\alpha}}{P_k}  (1+\psi(\gamma^{dl},\alpha)) } f_{{X}_k}^{D}(x) dx,
\end{equation}
with $\psi(\gamma^{dl},4)= \sqrt{\gamma^{dl}} \arctan{\sqrt{ \gamma^{dl}}}$ \cite{downlink}. The time, in seconds, necessary to complete the downlink transmission from the $k$th tier in coupled access is 
\begin{equation}
L^{dl}_{D,k}(\gamma^{dl})=\frac{B^O}{R_{D,k}^{dl}(\gamma^{dl})}.
\end{equation}

\subsubsection{Edge processing}

The computation servers in both tiers are assumed to have $M/M/1$ queuing system. With this model, the requests arrive according to a Poisson process with rate $\tau$ requests per second. The service rate are assumed to be independent and exponentially distributed with parameter $\mu$ requests per seconds. It is well known from queuing theory that the mean request delay of such servers is given by $1/(\mu-\tau)$. Motivated by this formula, we will derive a general expression for the computation time that captures both the tier in which the executions are performed and the load, in terms of the average number of associated users, of the serving BS as discussed next.

Let us first write the service rate $\mu_k$ in tier $k$ as 
\begin{equation} \label{mu}
\mu_k = \frac{F_k}{V},
\end{equation}
where $k=M$ for MBS and $k=S$ for SBS. It is noted here that the service rate is depending on the computation capability of the BS in tier $k$ that process the offloaded tasks. As mentioned in Section I, we realistically require $F_M \geq F_S$. 

Next, to calculate the request arrival rate $\tau_k$ in tier $k$, we define $d_k$ as the accumulated uplink rates for all the MUs that are connected to the same BS in tier $k$. We can then write
\begin{equation} \label{lam}
\tau_k = \frac{d_k}{B^I},
\end{equation}
with $k \in \{M,S\}$. To find an expression for $d_k$, we first observe that $d_k$ is depending on the number of MUs that are served by the $k$th BS in the UL. For a Voronoi cell, it is proved in \cite{area} that the average area of a Voronio cell in tier $k$ is $1/\lambda_k$. We then have the average number of served MUs as $(\lambda_u / \lambda_k) A_{kl}^D=N_{D,k}$ for $k \in \{M,S\} $. Building on these formulations, we can rewrite $d_k$ as
\begin{equation} \label{d}
d_k =  R_{D,k}^{ul} (\gamma^{ul}) \frac{\lambda_u}{\lambda_k}A_{kl}^{D}.
\end{equation}
By substituting (\ref{d}) in (\ref{lam}) and then using  (\ref{mu})-(\ref{lam}), we can have the general expression for the edge processing time at $k$th tier with coupled access as
\begin{equation}
L^{\rm exe}_{D,k} =\dfrac{1}{\mu_k - \tau_k} = \frac{1}{
\dfrac{F_k}{V}-\dfrac{ R_{D,k}^{ul}(\gamma^{ul})}{B^I}  N_{D,k} },
\end{equation}
where $k=M$ for MBS and $k=S$ for SBS. Clearly, $\mu_k > \tau_k $ is required for the stability of (19).

\subsection{Communication and Computation Model with Decoupled Access}

\subsubsection{Uplink transmission} Similar to (9), the average uplink rate of a typical MU
associated with $k$th tier using UL-based or decoupled association is given by 
\begin{equation}
R^{ul}_{U,k}(\gamma^{ul}
)=\frac{W^{ul}}{N_{U,k}^{ul}} \log_2(1+\gamma^{ul}) \mathrm{P}_{U,k}^{ul}(\gamma^{ul}),
\end{equation}
where $N_{U,k}^{ul}=\lambda_u A_k^{ul} /\lambda_k$ is the average number of associated MUs to a BS in the $k$th tier for the UL connection with $A_k^{ul}$ being the association probability for the $k$th tier BS in UL which depends on the association probabilities (4)-(7) according to the relation $A_k^{ul}=\sum_{l \in \{M,S\}} A_{kl}^U $. For instance, the probability that MU is connected to SBS tier in UL with decoupled access association is simply $A_S^{ul}=A_{SS}^{U}+A_{SM}^{U}$, where $A_{SS}^{U}$ and $A_{SM}^{U}$ are given in (4) and (6), respectively. The coverage probability for this case is given by \cite{petar}
\begin{equation}
\mathrm{P}_{U,k}^{ul}(\gamma^{ul})= \int_{0}^{\infty} e^{-\frac{\gamma^{ul} \sigma^2 x^{\alpha}}{P_u}} e^{-\pi \tilde \lambda_u \psi(\gamma^{ul},\alpha) x^2 } f_{{X}_k}^{U}(x) dx,
\end{equation}
with $f_{{X}_k}^{U}(x)$ is the PDF of the distance between a typical MU and its serving BS for UL-based association which reads \cite{petar}
\begin{equation}
f_{{X}_k}^{U}(x)=\frac{2 \pi \lambda_k}{A_k^{ul}}x e^{-(\lambda_k+\lambda_j) \pi x^2}.
\end{equation}
The decoupled-enabled transmission time in UL is
\begin{equation}
L^{ul}_{U,k}(\gamma^{ul})=\frac{B^I}{R_{U,k}^{ ul}(\gamma^{ul})}.
\end{equation}

\subsubsection{Downlink transmission}
Similar to the uplink, we can write the average rate in bits per seconds for the typical MU  associated to the $l$th tier in the DL using UL-based association as
\begin{equation}
R^{dl}_{U,l}(\gamma^{dl})=\frac{W^{dl}}{N^{dl}_{U,l}} \log_2(1+\gamma^{dl}) \mathrm{P}_{U,l}^{dl}(\gamma^{dl}),
\end{equation}
where $N_{U,l}^{dl}=\lambda_u A_l^{dl} /\lambda_l$ is the average number of associated MUs to a BS in the $l$th tier for the DL connection with decoupled access and $A_l^{dl}$ being the association probability for the $l$th tier BS in DL which depends on the association probabilities (4)-(7) according to the relation $A_l^{dl}=\sum_{k \in \{M,S\}} A_{kl}^U $. For instance, the probability that MU is connected to SBS tier in DL with decoupled access  is simply $A_S^{dl}=A_{SS}^{U}+A_{MS}^{U}$, where $A_{SS}^{U}$ and $A_{MS}^{U}$ are given in (4) and (7), respectively. The coverage probability $\mathrm{P}_{U,l}^{dl}(\gamma^{dl})$ is obtained from (14) after substituting $A_{kl}^D$ by $A_l^{dl}$ in (11). The corresponding required transmission time in DL with UL-based association reads
\begin{equation}
L^{dl}_{U,l}(\gamma^{dl})=\frac{B^O}{R_{U,l}^{ dl}(\gamma^{dl})}.
\end{equation}

\subsubsection{Edge processing}

With UL-based association, i.e., when the MU is connected to different BSs in UL and DL, the offloaded application can be processed at either one of the associated BSs. Accordingly, we distinguish the following two cases for edge processing:

\textit{(i) UL Cloudlet processing time:} The time needed to process the offloaded task at the UL cloudlet or $k$th tier BS is written as
\begin{equation}
L^{\rm exe}_{U,k} = \frac{1}{
\dfrac{F_k}{V}-\dfrac{ R^{ul}_{U,k}(\gamma^{ul})}{B^I}  N_{U,k}^{ul} }.
\end{equation}
We emphasize here that the expression in (26) is used to calculate the execution time for any association case that can result from decoupled access, i.e., $k=M$ and $k=S$ in UL, since in both cases the processing takes place at UL cloudlet. Therefore, it is natural for (26) to have similar expression to the coupled access execution time in (19). This is a direct result from the fact that decoupling the DL transmission from UL has no effect on the processing time when computations are performed at the BS to which MU are connected in UL. However, with DL cloudlet processing, the execution time need more careful consideration as discussed next. 

\textit{(ii) DL Cloudlet processing time:} 
If  the processing  takes place at  the  DL cloudlet or $l$th tier BS,  the execution time is given by
\begin{equation}
L^{\rm exe}_{U,l} = \frac{1}{
\dfrac{F_l}{V}-\left(\dfrac{ R^{ul}_{U,l}(\gamma^{ul})}{B^I}  N_{U,l}^{ul} + \dfrac{ R^{ul}_{U,k}(\gamma^{ul})}{B^I}  \dfrac{\lambda_u}{\lambda_k}A_{SM}^U\right)}.
\end{equation}
Unlike (26), the above expression applies specifically to determine the execution time only when $l=M$. This is because with DL cloudlet processing, the MBS receives requests from both the $N_{U,l}^{ul}$ fraction of MUs that are associated to MBS in UL and also the requests from $A_{SM}^U$ fraction of MUs that are transferred to DL cloudlet via backhaul links. For $l=S$, since the requests arrive to UL cloudlet via the UL connections only from $A_{SS}^U$ fraction of MUs, therefore the execution time reads
\begin{equation}
L^{\rm exe}_{U,l} = \frac{1}{
\dfrac{F_l}{V}-\dfrac{ R^{ul}_{U,k}(\gamma^{ul})}{B^I}  \dfrac{\lambda_u}{\lambda_k}A_{SS}^U}.
\end{equation}
\subsection{Offloading Latency}
We next formulate the overall offloading latency experienced by the typical MU for both coupled and decoupled schemes. We conclude this section by presenting the average offloading latency.

\subsubsection{Offloading latency with coupled access}
The offloading latency  for the typical MU associated to a given SBS or MBS, in both UL and DL, is 
 \begin{equation}
L^{D}_{kl}(\gamma^{ul}, \gamma^{dl})=L^{ul}_{D,k}(\gamma^{ul}) + L_{D,k}^{\rm exe} + L^{dl}_{D,k}(\gamma^{dl}),
\end{equation}
where $k=l=M$ for MBS and $k=l=S$ for SBS.

When the MU is connected to different BSs in UL and DL, the offloaded application can be processed at either one of the associated BSs. Accordingly, the following two cases are identified for decoupled access.

\subsubsection{Offloading latency with decoupled access and UL cloudlet processing} 

When the user is associated to a $k$BS in UL and a $l$BS in DL with $k,l \in \{M,S\}$ and processing takes place at the cloudlet of the $k$BS, the overall offloading latency experienced by the MU is then given by
\begin{equation}
\dot L^{U}_{kl}(\gamma^{ul}, \gamma^{dl})=L^{ul}_{U,k}(\gamma^{ul}) +L_{U,k}^{\rm exe} + L^{bh}_k+ L^{dl}_{U,l}(\gamma^{dl}).
\end{equation}
Note that, in this case, the backhaul is used to transfer the output bits produced by the execution at the $k$BS with corresponding transmission delay of $L^{bh}_k=B^O/C^{bh}$.

\subsubsection{Offloading latency with decoupled access and DL cloudlet processing}
When the user is associated to a $k$BS in UL and a $l$BS in DL with $k,l \in \{M,S\}$ and   processing  takes place at the cloudlet of the $l$BS, the offloading latency is
\begin{equation}
\ddot L^{U}_{kl}(\gamma^{ul}, \gamma^{dl})=L^{ul}_{U,k}(\gamma^{ul}) + L^{bh}_l+L_{U,l}^{\rm exe} + L_{U,l}^{dl}(\gamma^{dl}).
\end{equation}
In this case, the backhaul carries the input bits that are to be processed by the $l$BS with communication time given by $L^{bh}_l=B^I/C^{bh}$.

\subsubsection{ Average Offloading Latency}
Given the association probabilities in Sec. II-A and the offloading latency in (29)-(31), the average offloading latency for coupled access reads
\begin{equation}
 {L}^D(\gamma^{ul}, \gamma^{dl})=\sum_{k \in \{M,S\}} A^D_{kl} L_{kl}^D(\gamma^{ul}, \gamma^{dl}).
\end{equation}
with $l=k$. For the decoupled access with UL cloudlet processing we write
\begin{equation}
\dot {L}^U(\gamma^{ul}, \gamma^{dl})=\sum_{k \\ \in \{M,S\}} \sum_{l \in \{M,S\}} A^U_{kl} \dot L_{kl}^U(\gamma^{ul}, \gamma^{dl}),
\end{equation}
and similarly for DL cloudlet processing we have
\begin{equation}
\ddot {L}^U(\gamma^{ul}, \gamma^{dl})=\sum_{k \in \{M,S\}} \sum_{l \in \{M,S\}} A^U_{kl} \ddot L_{kl}^U(\gamma^{ul}, \gamma^{dl}).
\end{equation}

\section{Numerical Results}
In this section, we present numerical results to compare  the    latency performance of the MEC-based offloading scheme, in which decoupled access is adopted, with the conventional coupled offloading. To this end, we consider a two-tier setting with the macro tier described by $\lambda_M=1$ BS per sq. km, $P_M=46$ dBm \cite{petar,downlink} and $F_M=4.5$ GHz \cite{intel45}, while the tier of SBSs is characterized by $\lambda_S=10$ BS per sq. km, $P_S=30$ dBm \cite{petar,downlink} and  $F_S=3.9$ GHz \cite{intel45}. The user density is $\lambda_u=25$ MU per sq. km with $P_u=20$ dBm \cite{petar}. The uplink and downlink bandwidth is $W^{ul}=W^{dl}=1.4 $ MHz \cite{bw} and also we set $\sigma^2=-120$ dBm  with $\alpha=4$. We select $B^I=4 B^O$ with $B^O=1$ kbits  and the required CPU cycles of the offloaded components is set to $V=2640 \times B^I$ CPU cycles \cite{Miettinen2010EE}. Unless otherwise stated, we select the backhaul capacity as $C^{bh}=10$ Mbits/s \cite{ali} and $\gamma^{ul}=\gamma^{dl}=-10$ dB.

\begin{figure}[!t] \label{fig4}
        \centering
        \includegraphics[width=\columnwidth,height=.4\textheight,keepaspectratio]{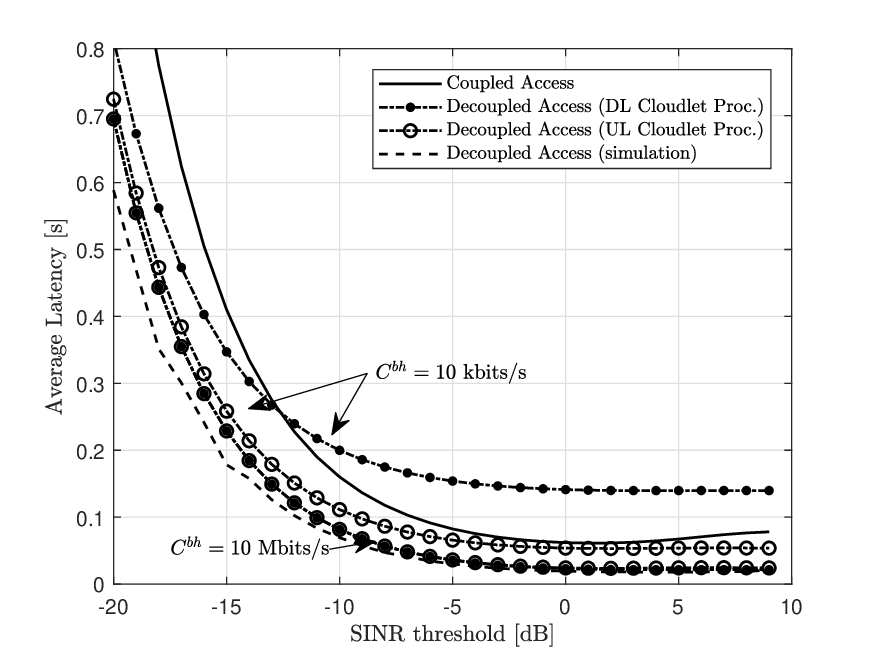}
        \caption{{\small{Average mobile offloading latency for coupled and decoupled access offloading schemes in (32)-(34) versus the target SINR threshold value $\gamma^{ul}=\gamma^{dl}$. }}}
        \label{model}
\end{figure}

Fig. 2 plots the offloading latency for the coupled access scheme in (32), marked as ``Coupled Access'' in Fig. 2, against the SINR thresholds in UL and DL which are assumed to be identical, i.e., $\gamma^{ul}=\gamma^{dl}$. Shown in the same figure, is the latency performance for decoupled access scheme given in (33) and (34) for UL and DL cloudlet processing, which are marked as ``Decoupled Access (UL Cloudlet Proc.)'' and ``Decoupled Access (DL Cloudlet Proc.)'', respectively  along with Monte Carlo simulations result for the decoupled access scheme with $C^{bh} =10$ Mbits/s. The  key  observation  here  is  that  the  performance of  the  decoupled access  is  strongly  limited by the backhaul capacity. For instance,  with  backhaul 10 Mbits/s the decoupled access scheme
starts to have a noticeable reduction in offloading latency. At SINR $\gamma^{ul}=\gamma^{dl}=-15$ dB, for  example,  the   decoupled access  scheme  attains $42\%$ latency reduction as compared to the coupled access scheme. The reason for this improvement, with the improved backhaul connections (e.g., fiber optic channel) between the two tiers, is the fact that the decoupled access scheme requires a certain fraction of MUs to decouple their associations between the two tiers, i.e., $k=S$ and $l=M$ for  $A_{SM}^U$ percentage of users since $A_{MS}^U =0$. This decoupling in UL/DL association brings two improvements to the performance of the decoupled scheme: first, an improved UL time for the $A_{SM}^U$ fraction of MUs due to the physical proximity, and hence better UL coverage, and also the higher density of SBSs tier compared to the tier of MBSs which implies more availability of radio resources, e.g., uplink bandwidth. Second, as a result of the first point, for the fraction of MUs which remain in coupled access to MBSs with UL-based association, i.e., $A_{MM}^U$, will now   benefit from both the more availability of resources and the enhanced coverage since the number of MUs coupled to  MBS in UL is decreased after decoupling the association of some MUs. These two gains combine to yield the latency saving of the decoupled-based offloading scheme as compared to the traditional coupled offloading. The slight increase in the average latency of coupled access scheme is due to higher arrival rate of requests as reflected by (19).

\section{Concluding Remarks}
The offloading of mobile computations with
independent association of  UL and DL transmissions is investigated by merging offloading decision and the MU-BS association into a single system perspective. The offloading scheme with decoupled association yields up to $42\%$ percent reduction in the total offloading latency compared to coupled access offloading.

\section*{Acknowledgement}
We thank professors O. Simeone, Dong Min and P. Popovski for suggesting the problem and part of the initial model.

\bibliographystyle{IEEEtran}
\bibliography{Al-Shuwaili_WCL2021-0771R1}

\end {document}